# Tuning turbine rotor design for very large wind farms

**Takafumi Nishino**[1,*] **and William Hunter**[2]

[1]Department of Engineering Science, University of Oxford, Oxford OX1 3PJ, United Kingdom

[2]Feldstr. 24, 85521 Ottobrunn, Germany

A new theoretical method is presented for future multi-scale aerodynamic optimisation of very large wind farms. The new method combines a recent two-scale coupled momentum analysis of ideal wind turbine arrays with the classical blade-element-momentum (BEM) theory for turbine rotor design, making it possible to explore some potentially important relationships between the design of rotors and their performance in a very large wind farm. The details of the original two-scale momentum model are described first, followed by the new coupling procedure with the classical BEM theory and some example solutions. The example solutions, obtained using a simplified but still realistic NREL S809 airfoil performance curve, illustrate how the aerodynamically optimal rotor design may change depending on the farm density. It is also shown that the peak power of the rotors designed optimally for a given farm (i.e. 'tuned' rotors) could be noticeably higher than that of the rotors designed for a different farm (i.e. 'untuned' rotors) even if the blade pitch angle is allowed to be adjusted optimally during the operation. The results presented are for ideal very large wind farms and a possible future extension of the present work for real large wind farms is also discussed briefly.

**Key words:** Actuator disc, Blade element, Momentum theory, Rotor design, Turbine arrangement, Wind farm optimisation

## 1. Introduction

The design of a large wind farm is a highly complex multi-disciplinary problem involving, among others, aerodynamic, structural, electrical, logistic and environmental considerations. As the goal of this multi-disciplinary design is not just a high total power output but also a low total cost and low environmental impact, there is no single optimal solution to the design of a large wind farm. Another important aspect of (especially aerodynamic) design of a large wind farm is its multi-scale nature, as highlighted in a recent statement of long-term research challenges issued by the European Academy of Wind Energy (van Kuik *et al.* 2016). The overall performance of a large wind farm depends on the characteristics of airflow across a wide range of length- and time-scales, making it very difficult to develop a comprehensive prediction model accounting for all relevant flow physics and their inter-scale interactions simultaneously, regardless of whether the modelling is experimental, computational or even theoretical.

Although the modelling of a large wind farm is truly challenging and needs further improvements in the coming years, there have already been many studies attempting the optimisation of large wind farms in recent years. For the optimisation of large wind farms there are a number of possible design parameters or variables, which may be broadly classified into the following: (i) parameters for the physical design of each wind turbine; (ii) parameters for the arrangement of wind turbines; and (iii)

---

* Email address for correspondence: takafumi.nishino@eng.ox.ac.uk

parameters for the operation of wind turbines/farm. Hence in general we may consider three different levels of wind farm optimisations[†] in terms of the types of parameters to be optimised:

Level 1: Optimising operating conditions only;

Level 2: Optimising turbine arrangement as well as operating conditions; and

Level 3: Optimising everything, including the design of each turbine.

Level 1 optimisation is essentially for people who have already built a wind farm, where it is usually no longer possible to change the physical design or arrangement of wind turbines but still possible to optimise operating conditions, such as the rotating speed and yaw angle of each turbine. For example, Fleming *et al.* (2014) have investigated potential benefits of redirecting turbine wakes by means of yaw-angle control, tilt-angle control and individual blade-pitch control; this type of optimisation may be classified as Level 1 optimisation. In contrast, Level 2 optimisation is essentially for people who have already finalised the design of turbines but have not built a farm yet. There are many recent studies on how the power of an entire farm could potentially be increased by changing the layout of turbine array (e.g. Meyers & Meneveau 2012; Wu & Porté-Agel 2013; Archer, Mirzaeisefat & Lee 2013; Stevens, Gayme & Meneveau 2014, 2016); these types of farm optimisations may be classified as Level 2 optimisation.

Whilst a number of studies have already been reported on Level 1 and Level 2 optimisation, there are only limited investigations of Level 3 optimisation in the literature. For example, Chamorro *et al.* (2014) have studied how the use of variable-sized wind turbines in a single farm could improve the overall performance of the farm. Similarly, Xie *et al.* (2017) have investigated potential benefits of collocating vertical-axis turbines with horizontal-axis turbines in a large wind farm. These studies are distinct from Level 2 optimisation in the sense that they consider the use of turbines with more than one specific design to improve the performance of a farm. However, these studies do not consider re-designing the details of each turbine, such as the shape of rotor blades, to make the turbine design particularly suitable for use in a large wind farm. To the authors' knowledge, no systematic study has been reported on how the performance of a large wind farm could be improved by modifying or 'tuning' the design of each turbine in the farm.

In this study we investigate potential benefits of tuning the aerodynamic design of horizontal-axis turbine rotors for use in a very large wind farm. This is motivated by recent theoretical studies (Nishino 2016; Zapata, Nishino & Delafin 2017) showing the possibility that the optimal local axial induction factor for turbines operating in a large wind farm could be significantly lower than that for an isolated turbine. This suggests that the power of a large farm could be improved by changing the physical design of turbines as well as their operating conditions. The theoretical model used in these recent studies is based on a rather simplified momentum conservation argument and should therefore be regarded as a rough approximation of a real farm (as explained further in the following sections). Nevertheless, considering the complex multi-scale nature of this Level 3 optimisation problem, it is worthwhile to start with such a basic theoretical model and extend it step-by-step. Specifically, here we combine this theoretical farm model with the classical blade-element-momentum (BEM) model to explore some basic mechanisms of possible inter-scale interactions between the optimal design of rotor blades and the layout (or 'density' in particular) of turbines in a very large wind farm.

In this paper we discuss the design of horizontal-axis turbines only. Another potentially interesting topic is the design of vertical-axis turbines optimised for large wind farms; however, this is outside the scope of the present paper.

---

[†] This classification was proposed by the first author in his keynote lecture at the 13[th] European Academy of Wind Energy (EAWE) PhD Seminar held at Cranfield University in the UK in September 2017.

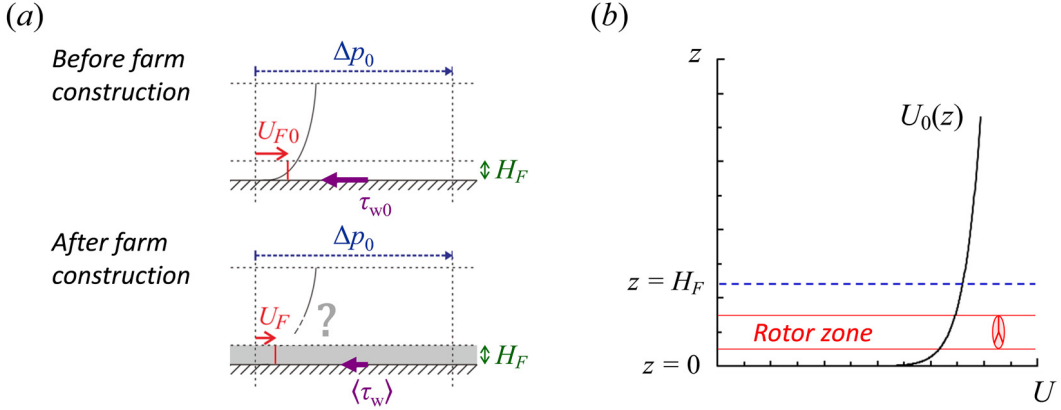

FIGURE 1. Schematic of the two-scale momentum model: (*a*) fully-developed boundary layers observed before and after farm construction; and (*b*) nominal farm-layer height $H_F$.

## 2. Theoretical model

### 2.1 Two-scale momentum model for a very large wind farm

The theoretical model used in this study is an extended version of the two-scale momentum model proposed by Nishino (2016); hence the original two-scale momentum model (hereafter referred to as the Nishino model) is described first in this subsection. Similarly to traditional 'top-down' models (e.g. Frandsen 1992; Emeis & Frandsen 1993; Calaf, Meneveau & Meyers 2010), the Nishino model considers momentum balances for a fully-developed boundary layer over a very large array of wind turbines (i.e. fully-developed wind farm canopy boundary layer); however, a major difference lies in the treatment of the boundary layer profile. In traditional top-down models, a fully-developed farm boundary layer profile is modelled usually as a combination of logarithmic functions; whereas in the Nishino model, the boundary layer profile is not modelled explicitly and only an average wind speed across a lower part of the boundary layer is modelled. The Nishino model may therefore be seen as a (very) low-order, quasi-one-dimensional flow model in contrast to other top-down models which are essentially two-dimensional flow models.

Figure 1 shows a schematic of the flow considered in the Nishino model. The flow is (in reality) turbulent and therefore the flow profiles depicted in this figure should be taken as temporally- and horizontally-averaged ones. As shown in figure 1*a*, the model considers a fully-developed boundary layer observed before constructing a wind farm (often referred to as the 'natural' or 'undisturbed' boundary layer) as well as that observed after constructing the farm. Although the natural boundary layer profile depicted here is a typical logarithmic one, we are allowed to consider different profiles and the only requirement is that the velocity gradient is zero or negligibly small at the upper end of the profile (so, for example, the profile may have a peak somewhere in the middle). As for the wind farm canopy, the Nishino model does not consider (at least explicitly) any differences in the type or pattern of turbine arrangement (aligned, staggered, etc.) but assumes that the turbines are arranged in a horizontally periodic manner (i.e. spaced evenly in the farm); hence a constant horizontal area, $S$, is allocated to each turbine, and the farm density, $\Lambda$, can be defined as

$$\Lambda = \frac{A}{S},$$

(2.1)

where $A$ is the rotor swept area of each turbine. (It should be noted that the original paper by Nishino (2016) uses $\lambda$ to describe the farm density, but the present paper uses $\Lambda$ instead since $\lambda$ will be used to describe the tip-speed ratio of the rotor.) A typical value of $\Lambda$ is in the range of 0.003 to 0.03.

A major assumption employed in this quasi-one-dimensional model is that the flow over the wind farm is driven by a constant streamwise pressure gradient (as indicated by $\Delta p_0$ in figure 1*a*). This may or may not be a good approximation of the real atmospheric flow over a large wind farm depending on the condition of the atmosphere and the size and geographical location of the farm, but suffices for the present study; see Appendix B for further discussion on this. (Although outside the scope of this study, another possible approach to model the flow over a large wind farm is to consider that the flow is driven by a constant geostrophic wind, and the relationship between the pressure-gradient forcing approach and the geostrophic-wind forcing approach has been discussed extensively by Calaf et al. (2010), for example.) By considering the momentum balance for a control volume extending from the bottom ($z = 0$, which corresponds to the ground or sea surface) to the top of this pressure-driven flow (where the velocity gradient is negligibly small) we obtain

$$\langle \tau_w \rangle S + T = \tau_{w0} S \,, \qquad (2.2)$$

where $\tau_{w0}$ and $\langle \tau_w \rangle$ are the (horizontally-averaged) shear stresses at the bottom of the control volume before and after farm construction, respectively, and $T$ is the rotor thrust of each turbine in the farm. The left-hand side of (2.2) represents the resisting force observed after farm construction (which is balanced by the driving force), whereas the right-hand side is that observed before farm construction (which is also balanced by the driving force). It should be noted that an additional term representing the resisting force due to support structures (turbine towers) could be added to the left-hand side of (2.2) as proposed by Ma and Nishino (2018); however, in the present study we neglect the effect of support structures for simplicity.

The equation (2.2) provides a relationship between $T$ and $\langle \tau_w \rangle$ but does not provide any details on how the turbines change the wind speed through the farm or through each rotor swept area. To relate $T$ and $\langle \tau_w \rangle$ with a representative wind speed through the farm (and then with the wind speed through each rotor), the Nishino model introduces a nominal farm-layer height, $H_F$, as illustrated in figure 1$b$. The height $H_F$ is based solely on the natural wind speed profile $U_0(z)$ in such a way that

$$U_{F0} = U_{T0} \,, \qquad (2.3)$$

where

$$U_{F0} = \frac{\int_0^{H_F} U_0 \, dz}{H_F} \quad \text{and} \quad U_{T0} = \frac{\int U_0 \, dA}{A} \qquad (2.4a \text{ and } b)$$

are the natural average wind speeds across the nominal farm layer and across the rotor swept area, respectively. As described by Nishino (2016), this nominal farm layer may be interpreted as a layer within which the flow is strongly affected by the turbines. However, this qualitative interpretation is for convenience only and the model does account for the possibility that the wind speed may change not only inside but also outside this nominal farm layer (since (2.2) is satisfied as long as the velocity gradient at the upper end of the entire flow is still negligibly small). With the above definition of the farm layer, we can also define the average wind speed across the farm layer, $U_F$, as well as that across the rotor swept area, $U_T$, as

$$U_F = \frac{\int \int_0^{H_F} U \, dz \, dS}{H_F S} \quad \text{and} \quad U_T = \frac{\int U \, dA}{A} \,, \qquad (2.5a \text{ and } b)$$

where $U$ represents the temporally-averaged (but not horizontally-averaged) streamwise velocity field observed after farm construction.

Using the three different wind speeds considered above, namely $U_{F0}(= U_{T0})$, $U_F$ and $U_T$, now we can define three different thrust coefficients for the turbine rotor:

$$C_T = \frac{T}{\frac{1}{2}\rho U_{F0}^2 A} \,, \qquad (2.6a)$$

$$C_T^* = \frac{T}{\frac{1}{2}\rho U_F^2 A} \,, \qquad (2.6b)$$

$$C_T' = \frac{T}{\frac{1}{2}\rho U_T^2 A} \,, \qquad (2.6c)$$

where $\rho$ is the density of air. Following the original paper by Nishino (2016), we refer to $C_T^*$ as the 'local' thrust coefficient in the present paper. Note that $C_T'$ may also be seen as another type of local thrust coefficient, but we refer to this as the 'resistance' coefficient (in analogy with that of a porous disc, which is often denoted by $k$ (e.g. Whelan, Graham & Peiró 2009) or $K$ (e.g. Nishino & Willden 2013) instead of $C_T'$). In the meanwhile, we can also define three different friction coefficients for the bottom boundary (ground or sea surface):

$$C_{f0} = \frac{\tau_{w0}}{\frac{1}{2}\rho U_{F0}^2}, \qquad (2.7a)$$

$$C_f = \frac{\langle \tau_w \rangle}{\frac{1}{2}\rho U_{F0}^2}, \qquad (2.7b)$$

$$C_f^* = \frac{\langle \tau_w \rangle}{\frac{1}{2}\rho U_F^2}. \qquad (2.7c)$$

Here $C_{f0}$ is the natural friction coefficient observed before farm construction, whereas $C_f^*$ is referred to as the 'local' or 'effective' friction coefficient observed after farm construction. If we assume that the flow through the nominal farm layer is self-similar before and after farm construction, then we would obtain $C_f^* = C_{f0}$ and hence $\langle \tau_w \rangle = (U_F/U_{F0})^2 \tau_{w0}$. However, this assumption may or may not be appropriate depending on how the turbines change the flow through the farm layer; therefore, the Nishino model introduces a more general representation of $\langle \tau_w \rangle$ as

$$\langle \tau_w \rangle = \beta^\gamma \tau_{w0}, \qquad (2.8)$$

where

$$\beta = \frac{U_F}{U_{F0}} \qquad (2.9)$$

is the ratio that indicates how much the farm-layer wind speed has decreased from the natural state, and the exponent $\gamma$ is the only model parameter that needs to be determined empirically in this quasi-one-dimensional flow model. Although the selection of an appropriate $\gamma$ value is not straightforward, the solution of this theoretical model (such as the power of each turbine) is usually less sensitive to the value of $\gamma$ than to other key parameters, as shown by Nishino (2016), who has also suggested that $\gamma = 2$ may be adopted to represent an ideal farm situation. By substituting (2.6b) and (2.8) into (2.2) for $T$ and $\langle \tau_w \rangle$, respectively, we obtain

$$C_T^* \frac{\Lambda}{C_{f0}} \beta^2 + \beta^\gamma - 1 = 0, \qquad (2.10)$$

where the combined parameter $\Lambda/C_{f0}$ is referred to as the 'effective' farm density.

Finally, the Nishino model uses the classical actuator disc momentum model as an approximation to estimate the local thrust coefficient $C_T^*$ in (2.10). Specifically, $C_T^*$ is modelled as

$$C_T^* = 4\alpha(1 - \alpha), \qquad (2.11)$$

where

$$\alpha = \frac{U_T}{U_F} \qquad (2.12)$$

is the ratio that indicates how much the farm-layer wind speed decreases locally at the turbine rotor, and hence $1 - \alpha$ may be referred to as a 'local' or 'effective' axial induction factor. Note that, since the actuator disc concept has been used to model $C_T^*$, it is implicitly assumed that the value of $\alpha$ is higher than approximately 0.6 (see, e.g. Hansen 2015). By substituting (2.11) into (2.10), we obtain the following two-scale coupled momentum balance equation:

$$4\alpha(1 - \alpha)\frac{\Lambda}{C_{f0}} \beta^2 + \beta^\gamma - 1 = 0, \qquad (2.13)$$

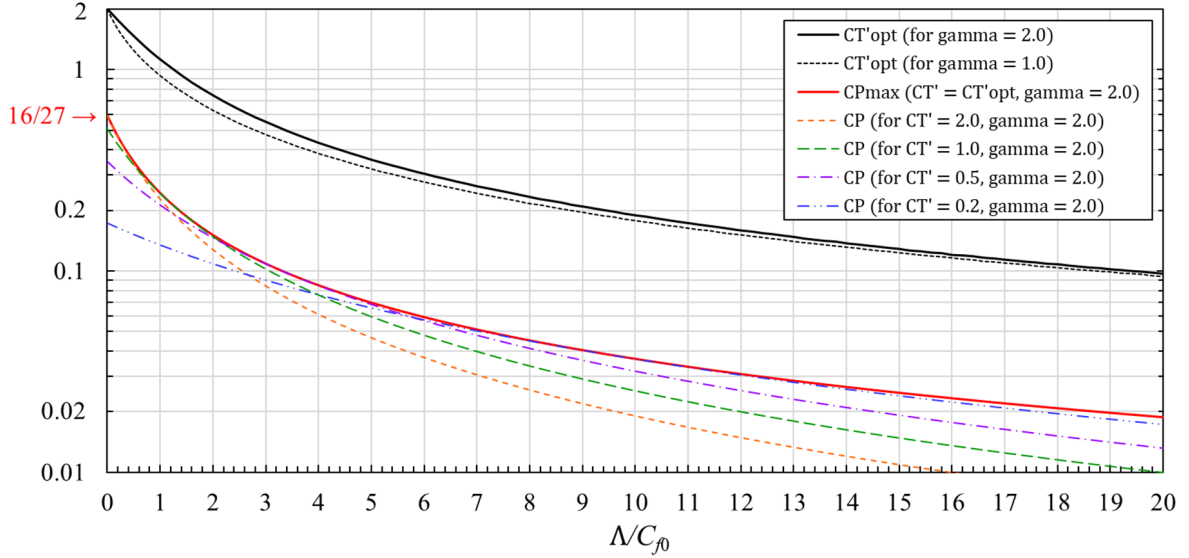

FIGURE 2. Optimal resistance coefficient ($C'_{T\mathrm{opt}}$) and maximum power coefficient ($C_{P\mathrm{max}}$) obtained from the original two-scale momentum model for very large wind farms (Nishino 2016).

which can be solved to obtain $\beta$ as a function of $\alpha$ for a given set of $\Lambda$, $C_{f0}$ and $\gamma$. Therefore, now we can calculate two different power coefficients of each turbine (actuator disc) in the farm:

$$C_P = \frac{TU_T}{\frac{1}{2}\rho U_{F0}^3 A} = 4\alpha^2(1-\alpha)\beta^3 \, , \qquad (2.14a)$$

$$C_P^* = \frac{TU_T}{\frac{1}{2}\rho U_F^3 A} = 4\alpha^2(1-\alpha) \, , \qquad (2.14b)$$

where $TU_T$ is the power of each actuator disc. In analogy with $C_T^*$ defined in (2.6b), $C_P^*$ is referred to as the 'local' power coefficient. We also obtain from (2.6b), (2.6c) and (2.11) that

$$C'_T = \frac{4(1-\alpha)}{\alpha} \, , \qquad (2.15)$$

and hence, we can also calculate $C_P$ as a function of $C'_T$ for a given set of $\Lambda$, $C_{f0}$ and $\gamma$. Of particular interest here is that the optimal $C'_T$ value ($C'_{T\mathrm{opt}}$) as well as the maximum $C_P$ value achieved ($C_{P\mathrm{max}}$) decreases exponentially as the effective farm density $\Lambda/C_{f0}$ increases, as can be seen in figure 2. For example, $C'_T = 2$ is optimal and yields $C_{P\mathrm{max}} = 16/27$ (i.e. the Betz limit) at $\Lambda/C_{f0} = 0$, but this high $C'_T$ value is not optimal and yields only about a half of $C_{P\mathrm{max}}$ at $\Lambda/C_{f0} = 10$, at which $C'_{T\mathrm{opt}}$ is about 0.2 instead of 2. Note that $\Lambda/C_{f0} = 10$ seems a realistic value for offshore wind farms, for which the value of $C_{f0}$ is typically of the order of 0.001. It should also be noted that these theoretical results are for ideal very large wind farms. For real large (but not infinitely large) wind farms, the value of $C_P$ may become higher than these theoretical values; see Appendix B for further details.

The validity of the Nishino model has been carefully investigated through comparisons with three-dimensional numerical simulations of a fully-developed flow over an infinitely large (periodic) array of porous discs (Zapata et al. 2017; Ghaisas, Ghate & Lele 2017; Dunstan, Murai & Nishino 2018). In summary, the key points of discussion in these validation studies are: (i) the use of the actuator disc model to estimate $C_T^*$ as in (2.11); (ii) the definition of the nominal farm-layer height based on (2.3); and (iii) how to determine the empirical parameter $\gamma$ in (2.8). For the use of (2.11), Zapata et al. (2017) have conducted a series of 3D Reynolds-averaged Navier-Stokes (RANS) simulations of a standard boundary-layer flow over aligned and staggered arrays of porous discs and found that the model agrees very well with the simulations of staggered discs (for a wide range of $\Lambda$ from 0.004 to 0.022). However, the model does not account for any differences in the pattern of disc arrangements and therefore tends to over-predict the power of aligned discs, suggesting that the use of (2.11) is appropriate only when there is no significant 'direct' wake effect, i.e. effect due to direct interference of the wake of the turbine immediately upstream. As for the use of (2.3), Dunstan et al. (2018) have conducted large-eddy simulations (LES) of more realistic boundary-layer flows with different atmospheric stability conditions and found that the theoretical model tends to over- and under-predict, respectively, the values of $\alpha$ and $\beta$ or vice versa, depending on the shape of the natural flow profile $U_0(z)$. However, the agreement of $C_P$ was still found to be satisfactory for all different flow profiles tested since the over-/under-predictions of $\alpha$ and $\beta$ tend to cancel each other out for $C_P$, suggesting that the use of (2.3) is acceptable as long as the prediction of $C_P$ is the main interest. With regard to the value of $\gamma$ in (2.8), Ghaisas et al. (2017) have reported that this may be modelled as a function of $\Lambda C_T'$; however, they were not able to assess the exact values of $\gamma$ since the resolution of their LES was found to be insufficient to do so. Their results also showed that $\gamma$ was around 2 for all cases (less than 2 for most cases). For simplicity, we adopt a fixed value of $\gamma = 2$ in the present study.

Before moving on to the extension of the Nishino model, we should also note the relevance of this work to some recent assessments of global wind energy resources, such as Adams and Keith (2013) and Miller et al. (2015), highlighting the importance of average wind speed reductions in large wind farms. As discussed by Nishino (2016), this theoretical model (with $\gamma = 2$) predicts that the value of $C_P \Lambda/C_{f0}$ (which is equivalent to the ratio of the power generated by the farm to the power dissipated naturally due to friction over the farm site before farm construction) increases asymptotically to the maximum value of about 0.385 as $\Lambda/C_{f0}$ is increased to a large value. This maximum power rate can also be obtained from a simpler (single-scale) momentum analysis, which was used by Miller et al. (2015) to predict the upper limit of large-scale wind power generation. We may therefore consider that the Nishino model describes, essentially, how the upper limit of wind power generation changes gradually from the one for isolated wind turbines (i.e. the Betz limit, $C_P \approx 0.593$) to the other for very large and dense wind farms ($C_P \Lambda/C_{f0} \approx 0.385$) as we increase the effective farm density.

### 2.2 Coupled BEM-farm-momentum (BEM-FM) model

The original two-scale momentum model described above is further extended in the present study by coupling the model with the classical BEM model with Prandtl's tip-loss factor (Glauert 1935; see also Hansen 2015). Figure 3 shows a conceptual diagram of the two-scale momentum, classical BEM and coupled BEM-farm-momentum (BEM-FM) models. In the coupled BEM-FM model, we use the BEM model (instead of the one-dimensional actuator disc momentum model) to calculate the local thrust coefficient $C_T^*$ in the farm-scale momentum balance equation (2.10) as well as the local power coefficient $C_P^*$ (note that these two local coefficients, $C_T^*$ and $C_P^*$, may be calculated without specifying the farm-scale parameters, namely $\Lambda$, $C_{f0}$ and $\gamma$, as will be shown below). The farm-scale momentum balance equation (2.10) is then solved to obtain $\beta$ as a function of $C_T^*$ (and therefore as a function of the rotor design and operating conditions specified in the BEM model) for a given set of $\Lambda$, $C_{f0}$ and $\gamma$. Eventually, the actual power coefficient $C_P$ (defined using the natural wind speed) is calculated from $C_P^*$ and $\beta$. Further details of the coupled model are described below.

First, we consider that the rotor swept area of each turbine is divided into $N$ annular elements (or annuli). Since the rotor thrust $T$ can be obtained from the summation of thrust on all $N$ elements, the local thrust coefficient $C_T^*$ can be expressed as

$$C_T^* = \frac{\sum_{i=1}^{N} \Delta T_i}{\frac{1}{2}\rho U_F^2 \pi R^2} , \qquad (2.16)$$

where $\Delta T_i$ is the thrust on the $i$-th annular element and $R$ is the rotor radius (so the rotor swept area $A = \pi R^2$). Following the blade-element theory, $\Delta T_i$ can be linked to the lift and drag forces on each blade element as

$$\Delta T_i = \frac{Z}{2}\rho V_i^2 c_i \Delta r_i \left(C_{L(i)} \cos \phi_i + C_{D(i)} \sin \phi_i\right) , \qquad (2.17)$$

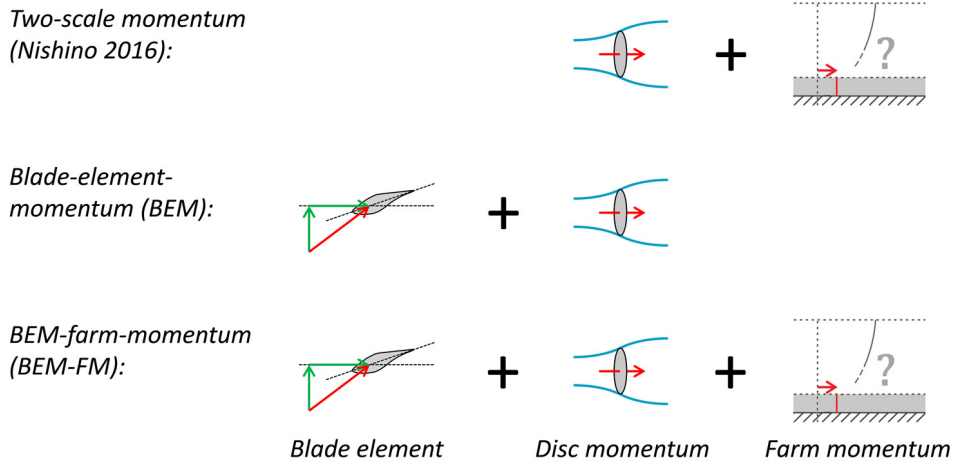

*Two-scale momentum (Nishino 2016):*

*Blade-element-momentum (BEM):*

*BEM-farm-momentum (BEM-FM):*

*Blade element*     *Disc momentum*     *Farm momentum*

FIGURE 3. Conceptual diagram of the two-scale momentum, blade-element-momentum (BEM) and BEM-farm-momentum (BEM-FM) models.

where $Z$ is the number of blades, $c_i$ and $\Delta r_i$ are the chord length and spanwise width of the $i$-th blade element, $C_{L(i)}$ and $C_{D(i)}$ are the lift and drag coefficients of the $i$-th blade element defined as

$$C_{L(i)} = \frac{L_i}{\frac{1}{2}\rho V_i^2 c_i \Delta r_i} , \qquad (2.18)$$

$$C_{D(i)} = \frac{D_i}{\frac{1}{2}\rho V_i^2 c_i \Delta r_i} , \qquad (2.19)$$

where $L_i$ and $D_i$ are the lift and drag forces on the $i$-th blade element, and

$$V_i^2 = \left( U_F(1 - a_i) \right)^2 + \left( \Omega r_i (1 + a_i') \right)^2 , \qquad (2.20)$$

$$\phi_i = \arcsin\left( \frac{U_F(1-a_i)}{V_i} \right) = \arccos\left( \frac{\Omega r_i(1+a_i')}{V_i} \right) = \arctan\left( \frac{U_F(1-a_i)}{\Omega r_i(1+a_i')} \right) , \qquad (2.21)$$

where $\Omega$ is the angular velocity of the rotor, $r_i$ is the radius of the $i$-th annular element, and $a_i$ and $a_i'$ are the axial and tangential induction factors, respectively. It should be noted that $U_F$ has been used here as an approximation of the wind speed upstream of the rotor; this is in analogy with the original two-scale momentum model (Nishino 2016) and, as described earlier, this approximation seems to be valid only when there is no significant 'direct' wake effect between neighbouring turbines. With this approximation, the tip speed ratio of the rotor is also defined as

$$\lambda = \frac{\Omega R}{U_F} , \qquad (2.22)$$

which is an operating parameter of the rotor. It should also be noted that the values of $C_{L(i)}$ and $C_{D(i)}$ depend mainly on the airfoil shape used for the cross-section of the blade and its local angle of attack (AoA), the latter of which is calculated as

$$\text{AoA}_i = \phi_i - \left( \theta_p + \theta_i \right) , \qquad (2.23)$$

where $\theta_p + \theta_i$ is the local angle between the airfoil chord and the rotor plane; $\theta_p$ is the angle due to the pitch of the entire blade, whereas $\theta_i$ is the angle due to the twist of the blade. Note that $\theta_p$ is an operating parameter of the rotor (constant for all $N$ blade elements), whereas $\theta_i$ is a design parameter of the rotor (function of the radial coordinate).

Similarly to $C_T^*$ in (2.16), the local power coefficient $C_P^*$ can also be expressed using a summation of contributions from $N$ annular elements:

$$C_P^* = \frac{\sum_{i=1}^N \Delta Q_i \Omega}{\frac{1}{2}\rho U_F^3 \pi R^2} \ , \tag{2.24}$$

where

$$\Delta Q_i = \frac{Z}{2}\rho V_i^2 c_i \Delta r_i \big(C_{L(i)}\sin\phi_i - C_{D(i)}\cos\phi_i\big) r_i \tag{2.25}$$

is the torque on the $i$-th element. Again, $V_i$ and $\phi_i$ depend on $a_i$ and $a_i'$ as in (2.20) and (2.21); hence $a_i$ and $a_i'$ need to be determined to obtain $C_T^*$ and $C_P^*$ for a given set of rotor design parameters (such as $Z$, $c_i$ and $\theta_i$) and operating parameters ($\lambda$ and $\theta_\mathrm{p}$).

To determine $a_i$ and $a_i'$, the classical actuator disc momentum theory can be adopted (again using $U_F$ as an approximation of the wind speed upstream of the rotor), giving the following relationships for each annular element (see, e.g. Hansen 2015):

$$\Delta T_i = 4a_i(1-a_i)F_i \rho U_F^2 \pi r_i \Delta r_i \ , \tag{2.26}$$

$$\Delta Q_i = 4a_i'(1-a_i)F_i \rho U_F \Omega \pi r_i^3 \Delta r_i \ , \tag{2.27}$$

where $F_i$ is the value of Prandtl's tip loss factor (Glauert 1935) for the $i$-th element:

$$F_i = \frac{2}{\pi}\arccos\left(\exp\left(-\frac{Z(R-r_i)}{2r_i\sin\phi_i}\right)\right). \tag{2.28}$$

Since $\Delta T_i$ and $\Delta Q_i$ in (2.26) and (2.27), respectively, should agree with those in (2.17) and (2.25), the values of $a_i$ and $a_i'$ that satisfy these equations can be calculated in an iterative manner. Note that the above equations include the 'upstream' wind speed $U_F$, which depends on the farm-scale momentum balance, but this unknown variable can be eliminated using (2.21) and (2.22) after dividing both sides of (2.17), (2.25), (2.26) and (2.27) by $U_F^2$; therefore, we may consider that the above BEM calculation is independent of the farm-scale problem (provided that $C_{L(i)}$ and $C_{D(i)}$ are insensitive to changes in the Reynolds number due to changes in $U_F$). It should also be noted that we do not employ Glauert's empirical correction to the calculation of $\Delta T_i$ (e.g. Hansen 2015) in the present study for simplicity.

Once $a_i$ and $a_i'$ have been determined, we can calculate $C_T^*$ and $C_P^*$ as

$$C_T^* = \frac{\sum_{i=1}^N \Delta T_i}{\frac{1}{2}\rho U_F^2 \pi R^2} = \frac{2}{R^2}\sum_{i=1}^N 4a_i(1-a_i)F_i r_i \Delta r_i \ , \tag{2.29}$$

$$C_P^* = \frac{\sum_{i=1}^N \Delta Q_i \Omega}{\frac{1}{2}\rho U_F^3 \pi R^2} = \frac{2\lambda^2}{R^4}\sum_{i=1}^N 4a_i'(1-a_i)F_i r_i^3 \Delta r_i \ , \tag{2.30}$$

for a given rotor design and operating conditions. Then the value of $C_T^*$ can be substituted into (2.10) to obtain $\beta$ for a given set of $\Lambda$, $C_{f0}$ and $\gamma$. Finally, the thrust and power coefficients of each rotor in the farm (defined using the natural wind speed $U_{F0}$ instead of $U_F$) can be calculated as

$$C_T = \frac{\sum_{i=1}^N \Delta T_i}{\frac{1}{2}\rho U_{F0}^2 \pi R^2} = \beta^2 C_T^* \ , \tag{2.31}$$

$$C_P = \frac{\sum_{i=1}^N \Delta Q_i \Omega}{\frac{1}{2}\rho U_{F0}^3 \pi R^2} = \beta^3 C_P^* \ , \tag{2.32}$$

for a given set of rotor design, operating and farm-scale parameters.

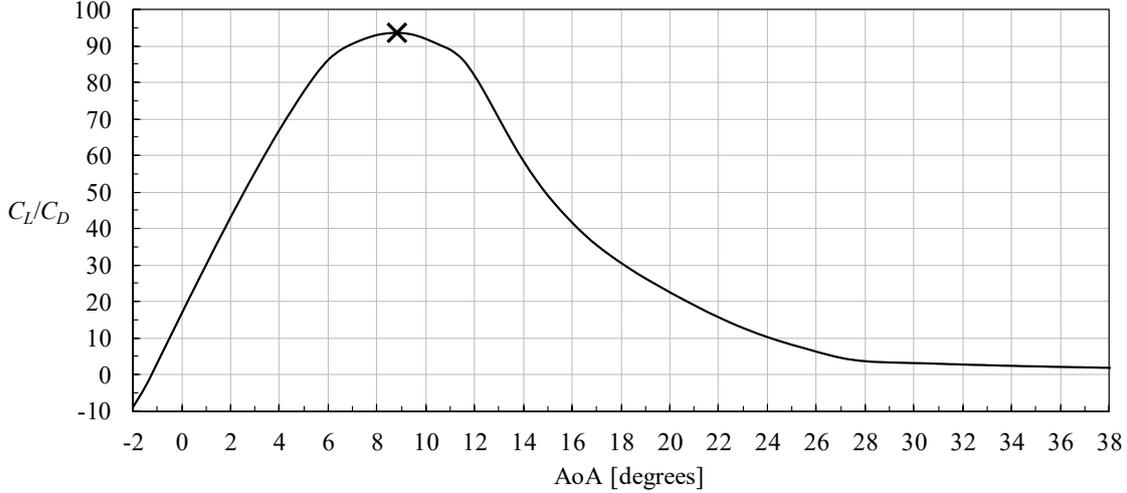

FIGURE 4. The $C_L/C_D$ curve used in the present study. The cross mark indicates the peak of the curve: $(C_L/C_D)_{max} = 93.71$ at AoA = 8.78 degrees.

## 3. Example solutions

Some example solutions of the coupled BEM-FM model are presented below to illustrate how the model may be used to explore potential benefits of tuning horizontal-axis rotor design for very large wind farms. First, we consider a special case with the effective farm density $\Lambda/C_{f0}$ being zero (which means that each turbine in the farm is completely isolated) and find an aerodynamically optimal rotor design under a certain set of restrictions. Secondly, we investigate how this optimal rotor design may change with $\Lambda/C_{f0}$, i.e. how the rotor design may be 'tuned' for different farm densities. Finally, we compare the performance of tuned and untuned rotors at different farm densities to explore the benefits of tuning. In particular, it is demonstrated that the power of untuned rotors cannot be as high as that of tuned rotors even if their operating conditions ($\lambda$ and $\theta_p$) are optimally adjusted.

For this demonstration, we employ a simplified (but still realistic) performance curve of the NREL S809 airfoil to determine force coefficients of each blade element as a function of AoA, as shown in figure 4. This curve has been created by blending two sets of solutions of the two-dimensional airfoil analysis code XFOIL (Drela 1989) for a range of AoA between 0 and 30 degrees (one is with forced transition near the leading edge, from 0 to 11 degrees, and the other is with free transition, from 11 to 30 degrees; both are for a chord Reynolds number of $10^7$), extending the data using AirfoilPrep (Ning 2013) for a wider range of AoA, and finally smoothing the data for the entire AoA range. The optimal AoA is 8.78 degrees, at which we observe the maximum $C_L/C_D$ of 93.71 (or the minimum $C_D/C_L$ of 0.01067). It should be noted that, in reality, the lift and drag coefficients of an airfoil may depend not only on the airfoil shape and AoA but also on other physical factors, such as the freestream turbulence level and the variation of local Reynolds number for each blade element (defined based on $V_i$ and $c_i$). However, we do not consider the influence of these factors in the present study for simplicity.

### 3.1 Optimal rotor design for isolated turbines ($\Lambda/C_{f0} = 0$)

First, we consider optimising the design of rotors for the case with $\Lambda/C_{f0} = 0$. In this special case, we obtain $\beta = 1$ from (2.10), meaning that the coupled BEM-FM model returns to the classical BEM model (as the turbines are located so far apart that they do not affect the average wind speed through the nominal farm layer). Unless specified otherwise, the number of blades for each rotor is fixed at three ($Z = 3$) during the process of optimisation. It is also assumed that the length of each blade is 90% of the rotor radius; hence the blade root and tip are positioned at $r/R = 0.1$ and 1, respectively. Even with these restrictions, there are still multiple ways to obtain the aerodynamically optimal rotor design, i.e. the set of radial variations of the blade chord length $c$ and twist angle $\theta$ that is required to maximise the power coefficient $C_P$ in (2.32). Here we consider finding an optimal rotor design for a given tip speed ratio $\lambda$ first, using the following BEM procedure:

Step 1: Assigning an arbitrary value to $a_i$ (within the range of 0 to 0.4) and an initial value to $a'_i$ (e.g. $a'_i = 0$) for each element;

Step 2: Calculating $\phi_i$ from (2.21) for each element;

Step 3: Calculating $\theta_i$ from (2.23) for each element, considering that $\mathrm{AoA}_i$ stays at the optimal value (8.78 deg. as shown in figure 4) and $\theta_P$ is zero for the optimal rotor sought here;

Step 4: Calculating $F_i$ from (2.28) for each element;

Step 5: Calculating the value of $\Delta T_i/\rho U_F^2$ from (2.26) for each element;

Step 6: Calculating $c_i$ from (2.17) for each element;

Step 7: Calculating the value of $\Delta Q_i/\rho U_F^2$ from (2.25) for each element;

Step 8: Calculating $a'_i$ from (2.27) for each element;

Step 9: Repeating Step 2 to Step 8 until the solutions converge for each element;

Step 10: Calculating $C_P^*$ (which is identical to $C_P$ in this case since $\beta = 1$) from (2.30);

Step 11: Repeating Step 1 to Step 10 to find the optimal $a_i$ that maximises $C_P$.

Once the set of solutions yielding the highest $C_P$ (for a given tip speed ratio $\lambda$) has been obtained, the above procedure is repeated for a range of $\lambda$ (between 1 and 10 in this study) to finally find the rotor design optimised for the optimal tip speed ratio for $\Lambda/C_{f0} = 0$. Note that the axial induction factor $a_i$ (not $a_i F_i$) assigned in this rotor optimisation procedure can be assumed to be uniform in the radial direction. Strictly speaking, the optimal rotor design/performance may improve a little further by allowing non-uniform $a_i$ during the process of optimisation and, for the case with $\Lambda/C_{f0} = 0$, this could be easily implemented by searching for the optimal $a_i$ that maximises the value of $\Delta Q_i/\rho U_F^2$ for each element separately instead (as this is equivalent to maximising $C_P$ when $\beta = 1$). However, the improvements due to non-uniform $a_i$ tend to be negligibly small; for example, the increase in the highest $C_P$ is less than 0.5% at all tip speed ratios investigated and less than 0.1% at the optimal tip speed ratio. Therefore, we assume uniform $a_i$ when optimising the design of rotors in this study (to simplify the optimisation process, especially for the case with $\Lambda/C_{f0} > 0$ to be presented later).

Figure 5 shows the performance of optimally designed rotors for $\Lambda/C_{f0} = 0$; each circle represents the performance of a (sub-)optimal rotor (optimised for a given $\lambda$, with $Z = 3$ and $C_D/C_L = 0.01067$), whereas the two dashed lines show how the optimal rotor (optimised for the optimal $\lambda$, which is 8.42 in this case as indicated by the cross mark) performs at different $\lambda$ with pitch control (i.e. with the pitch angle $\theta_P$ being adjusted to maximise $C_P$) and without pitch control (i.e. with $\theta_P$ being fixed at zero), respectively. Also presented here for comparison is the performance of (sub-)optimal rotors with different airfoil performance ($C_D/C_L = 0$ and 0.02, respectively, at their optimal AoA) and with an infinite number of blades ($Z = \infty$, corresponding to an ideal actuator disc with rotation) as well as the Betz limit. It can be confirmed that the optimal $\lambda$ increases with the optimal airfoil performance; $\lambda \approx 6$ at $C_D/C_L = 0.02$ increases to $\lambda = 8.42$ at $C_D/C_L = 0.01067$ and this goes up to infinity as $C_D/C_L$ approaches to zero. The rotor performance itself also increases with the airfoil performance, and this increases further with the number of blades (as the tip loss factor $F$ in (2.28) approaches to unity as $Z$ approaches to infinity). The performance of the optimal rotor operating at a non-optimal $\lambda$ is always lower than that of the rotor optimised specifically for that $\lambda$. Adjusting optimally the pitch angle $\theta_P$ may help improving the performance of the optimal rotor at a non-optimal $\lambda$, but this improvement is smaller than what is achieved by optimising the rotor design for that $\lambda$.

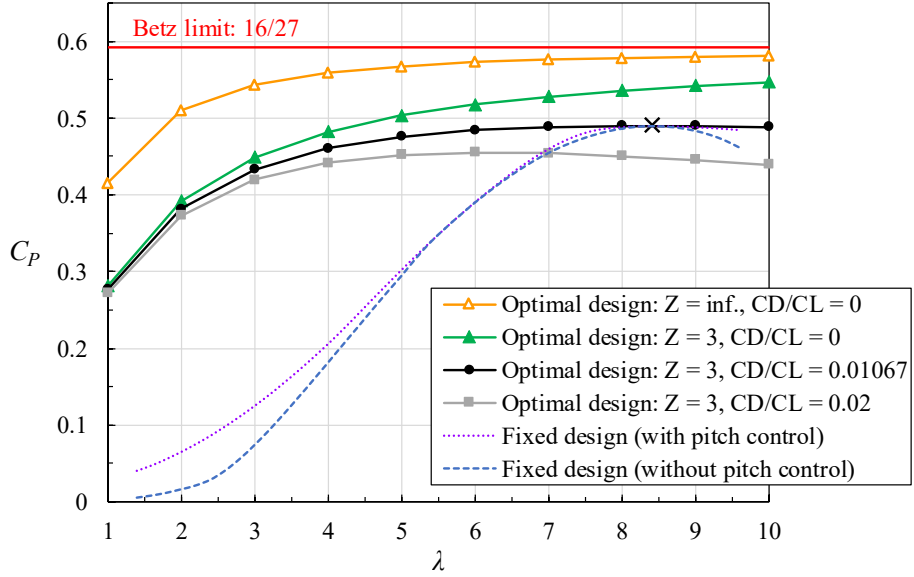

FIGURE 5. Comparison of the performance of optimally designed rotors for the case of isolated turbines ($\Lambda/C_{f0} = 0$). Each symbol point shows the performance of a rotor designed optimally for a given $\lambda$, whereas the dashed lines show how the rotor designed optimally for the optimal $\lambda$ (indicated by the cross mark) performs at different $\lambda$ with and without pitch control.

### 3.2 Tuned rotor design for turbines in a very large wind farm

Next, we consider optimising or 'tuning' the rotor design for different effective farm density cases (hereafter we shall use the word 'tuned' to refer specifically to the rotor design optimised for a given effective farm density). To demonstrate how the tuned rotor design tends to change with the effective farm density, here we consider six different effective farm densities: $\Lambda/C_{f0} = 0.2$, 1, 2, 5, 10 and 20. The procedure required to calculate the tuned rotor design is similar to that for $\Lambda/C_{f0} = 0$ presented earlier. The difference is that, for $\Lambda/C_{f0} > 0$, the farm-scale momentum balance equation (2.10) needs to be additionally solved to obtain $\beta$ and thereby $C_P$; therefore, Step 10 and Step 11 described earlier need to be replaced by the following steps:

Step 10: Calculating $C_T^*$ from (2.29);

Step 11: Calculating $\beta$ from (2.10);

Step 12: Calculating $C_P$ from (2.32);

Step 13: Repeating Step 1 to Step 12 to find the optimal $a_i$ that maximises $C_P$.

As noted earlier, the axial induction factor $a_i$ is assumed to be uniform in the radial direction during the rotor design process (but not when calculating the performance of an already designed rotor, for which $a_i$ is allowed to be non-uniform). Again, the above procedure is to calculate an optimal rotor design for a given $\lambda$ and is therefore repeated for a range of $\lambda$ to find the optimal (tuned) rotor design with the optimal $\lambda$.

Figure 6 shows the performance of optimally designed rotors for all six $\Lambda/C_{f0}$ values investigated. Again, each circle represents a (sub-)optimal rotor (with $Z = 3$ and $C_D/C_L = 0.01067$). Also shown for comparison are the theoretical $C_{P\max}$ values obtained from the original two-scale momentum model of Nishino (2016). It can be seen that the optimal $\lambda$ (indicated by the cross mark in each sub-figure) tends to decrease as $\Lambda/C_{f0}$ increases; this is essentially because the optimal rotor resistance decreases as $\Lambda/C_{f0}$ increases (as presented earlier in figure 2). Similarly to the previous case with $\Lambda/C_{f0} = 0$, at each $\Lambda/C_{f0}$, the optimal rotor operating at a non-optimal $\lambda$ always yields a lower $C_P$ compared to the corresponding (sub-)optimal rotor designed for that $\lambda$, regardless of whether or not the optimal pitch control is applied.

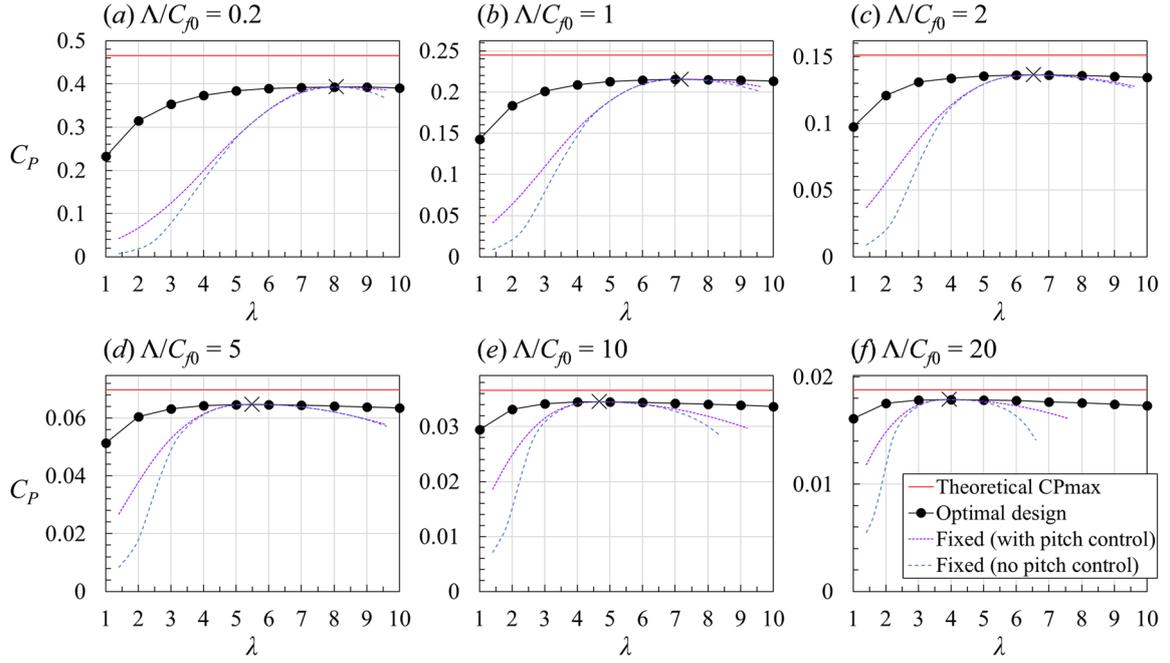

FIGURE 6. Performance of optimally designed rotors for different effective farm density cases; meanings of the symbols and dashed lines are the same as in figure 5.

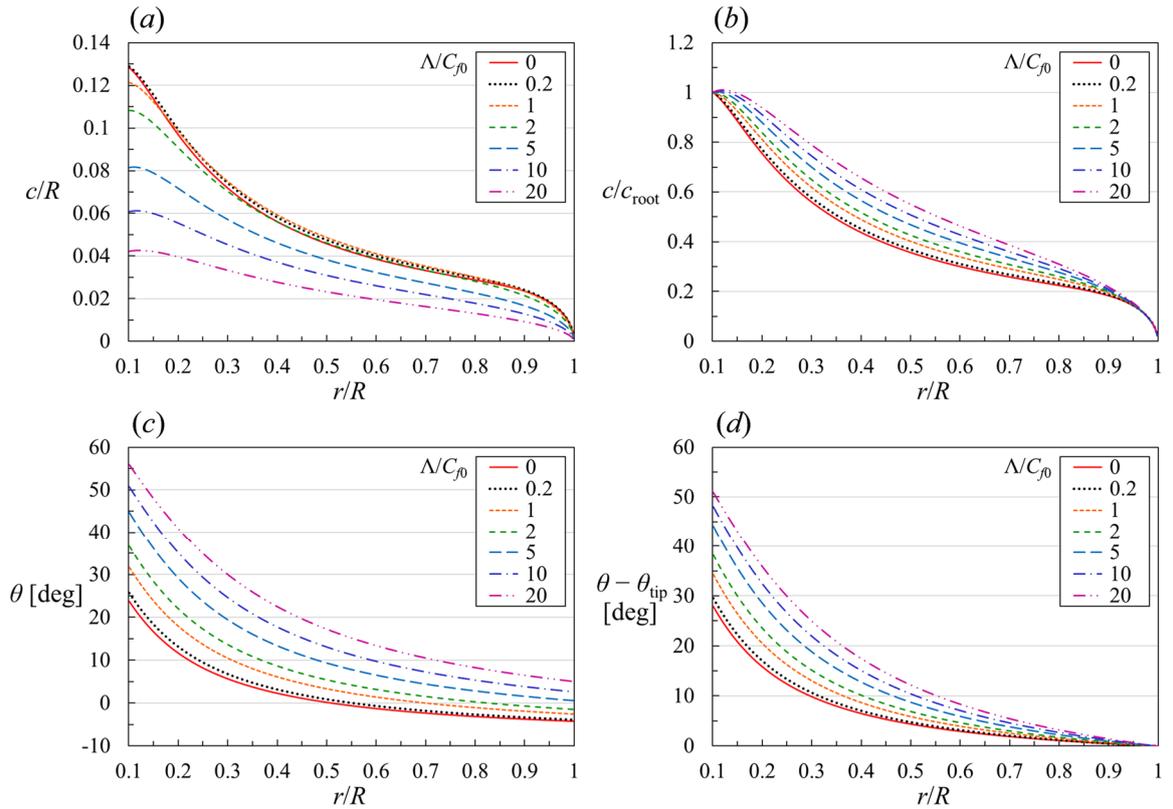

FIGURE 7. Tuned rotor design for different effective farm densities: ($a$, $b$) radial variation of the blade chord length, where $c_{\text{root}}$ is the chord length at $r/R = 0.1$; ($c$, $d$) radial variation of the blade twist angle, where $\theta_{\text{tip}}$ is the twist angle at $r/R = 1$.

Figure 7 compares the tuned rotor designs obtained, namely the radial variations of the blade chord length and twist angle, for the six different wind farm cases ($\Lambda/C_{f0}$ = 0.2, 1, 2, 5, 10 and 20) and the case of isolated turbines ($\Lambda/C_{f0}$ = 0). One of the major effects of the effective farm density $\Lambda/C_{f0}$ is that the optimal chord length tends to decrease across the entire blade span as $\Lambda/C_{f0}$ increases (except for $0 < \Lambda/C_{f0} < 1$, where the optimal chord length does not change substantially, cf. figure 7$a$). This trend, that the optimal rotor solidity tends to decrease as the farm density increases, may be predicted from the original two-scale momentum model showing that the optimal rotor resistance decreases as the farm density increases. However, as can be seen from figure 7$b$, the 'normalised' profiles of the optimal chord length for different $\Lambda/C_{f0}$ cases (normalised by the optimal root chord length $c_{\text{root}}$ for each case) do not collapse into a single profile, indicating that the coupled BEM-FM approach may indeed be helpful (more helpful than the original two-scale momentum model) to tune the design of rotors for very large wind farms. Another major effect of $\Lambda/C_{f0}$ is that on the variation of the optimal blade twist angle. It can be seen from figures 7$c$ and 7$d$ that the tuned rotor blades become more and more twisted as $\Lambda/C_{f0}$ increases.

### 3.3 *Performance comparison of tuned and untuned rotors*

Now we compare the performance of 'tuned' and 'untuned' rotors in a very large wind farm with various effective farm densities (here we use the word 'untuned' to refer to the rotor design that has been optimised for a different $\Lambda/C_{f0}$ value from that of the wind farm of interest). Specifically, we compare the performance of five different rotors: Rotor-0, Rotor-2, Rotor-5, Rotor-10 and Rotor-20, which are the rotors optimised for $\Lambda/C_{f0}$ = 0, 2, 5, 10 and 20, respectively.

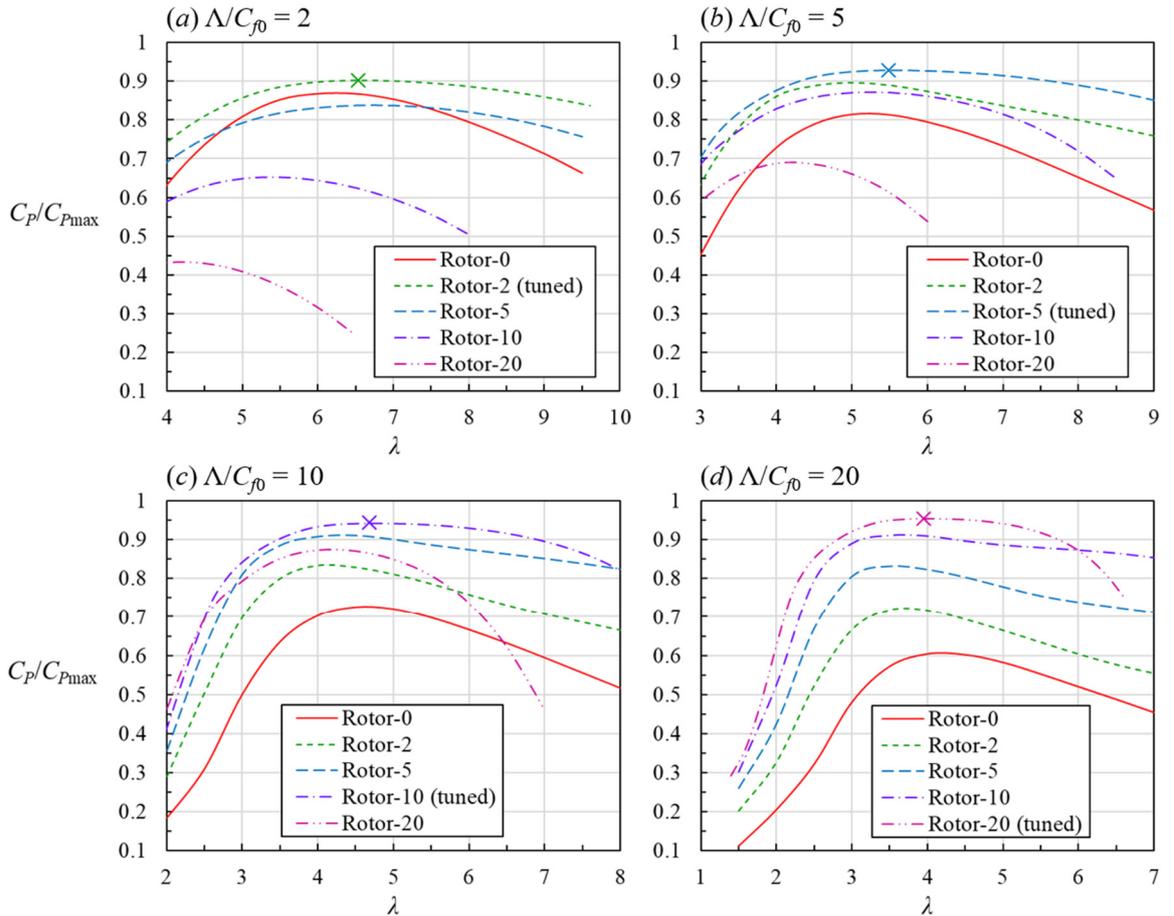

FIGURE 8. Comparison of the performance of tuned and untuned rotors operating without pitch control ($\theta_{\text{p}} = 0$) in a very large wind farm with different effective farm densities.

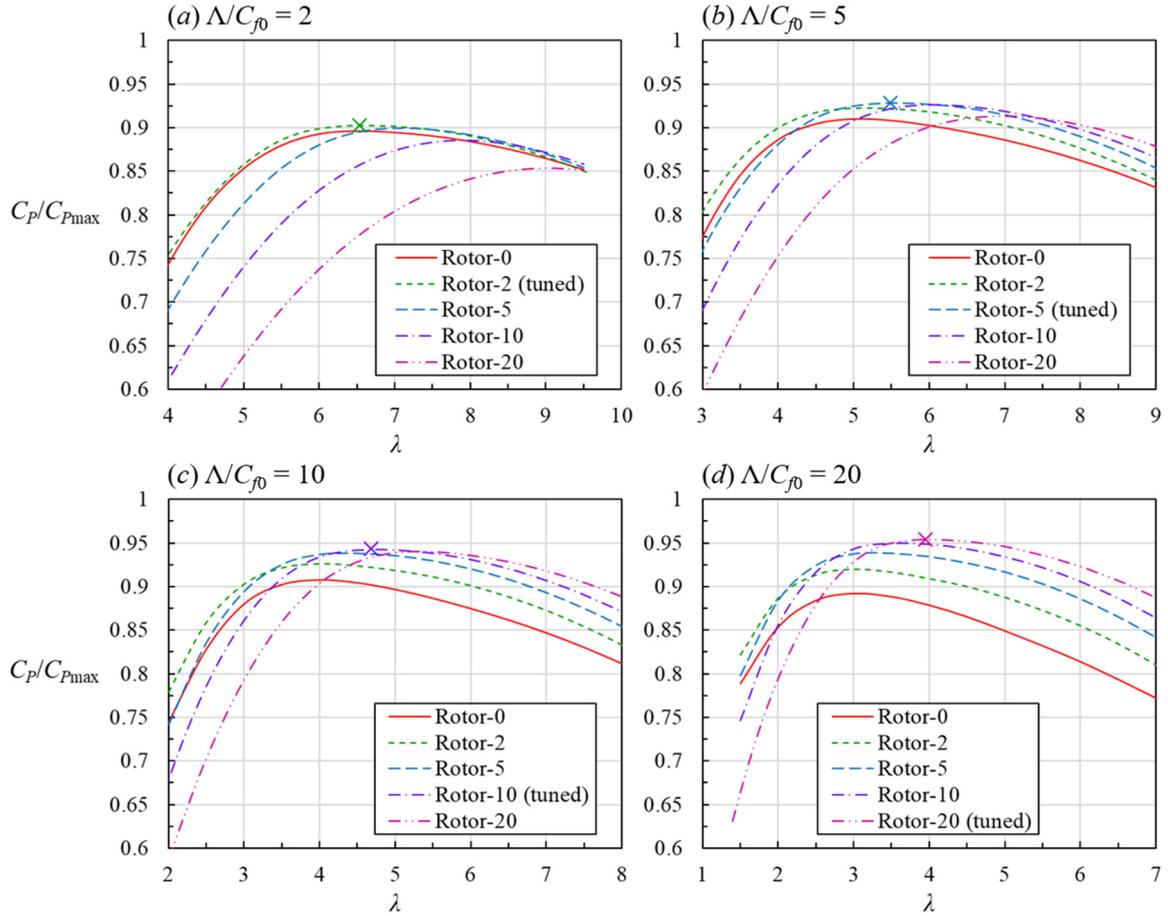

FIGURE 9. Comparison of the performance of tuned and untuned rotors operating with pitch control (i.e. $\theta_p$ adjusted optimally) in a very large wind farm with different effective farm densities.

Figure 8 shows the performance curves of the five rotors operating without pitch control (i.e. with $\theta_P = 0$) in a farm with $\Lambda/C_{f0} = 2$, 5, 10 and 20. Note that we have one tuned rotor and four untuned rotors for each $\Lambda/C_{f0}$ case, and the performance curve of the tuned rotor is the same as that presented earlier in figure 6 (although here $C_P$ has been normalised by the theoretical $C_{P\max}$ for each $\Lambda/C_{f0}$). It can be seen clearly that the performance difference between the tuned rotor and an untuned rotor can be significant when the pitch control is not applied. For example, in a 'dense' farm with $\Lambda/C_{f0} = 20$, the peak $C_P$ of the tuned rotor (Rotor-20) is about 95% of the theoretical $C_{P\max}$, whereas that of the rotor optimised for isolated turbines (Rotor-0) is only about 60% of the theoretical $C_{P\max}$. It can also be seen that the tuned rotor tends to maintain a high $C_P$ value for a wider range of the tip speed ratio compared to untuned rotors.

The above performance difference between the tuned and untuned rotors, however, becomes much smaller if the blade pitch angle $\theta_P$ is allowed to be adjusted optimally. Figure 9 compares the power curves of the five rotors operating with optimal pitch control, again in a farm with $\Lambda/C_{f0} = 2$, 5, 10 and 20. It can be seen that, for each $\Lambda/C_{f0}$ case, the tuned rotor still yields the highest $C_P$ value, but untuned rotors may also perform fairly well due to the pitch control. For example, at $\Lambda/C_{f0} = 20$, the peak $C_P$ of Rotor-0 increases up to 89% of the theoretical $C_{P\max}$ because of the pitch control, whilst that of the tuned rotor (Rotor-20) stays at 95% of the theoretical $C_{P\max}$ since the tuned rotor has been designed to yield the highest possible $C_P$ with $\theta_P = 0$, i.e. this highest $C_P$ cannot be increased further by changing $\theta_P$.

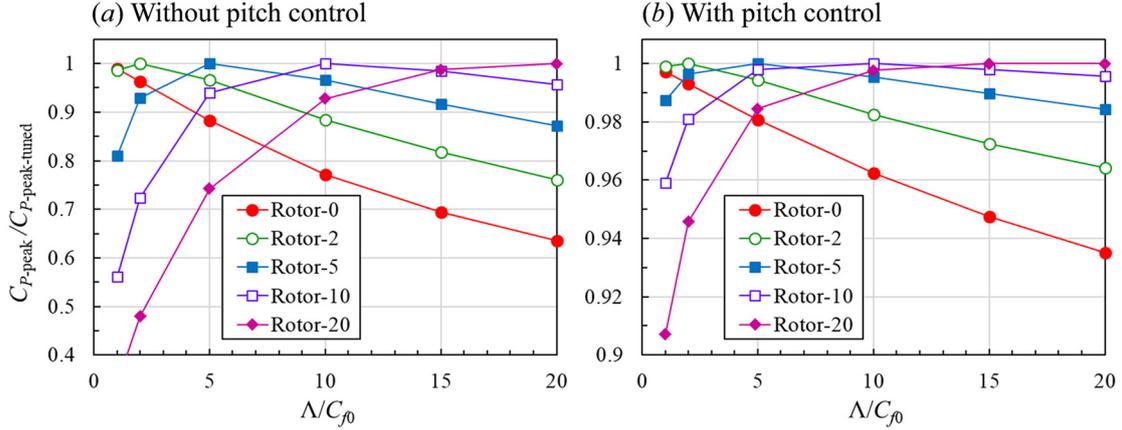

FIGURE 10. Comparison of the peak performance of tuned and untuned rotors at different effective farm densities: (*a*) without pitch control; and (*b*) with pitch control.

The trend of the peak performance of the five rotors can be seen more clearly in figure 10, which compares the peak $C_P$ of the five rotors (normalised by that of the tuned rotor at each $\Lambda/C_{f0}$) with and without pitch control. Note that, although not shown for brevity, here we have additionally calculated the tuned rotor design for $\Lambda/C_{f0} = 15$ as well for the normalisation of $C_P$ values at that $\Lambda/C_{f0}$. It can be seen that the overall trend is similar between the cases with and without pitch control, though the performance differences between the five rotors are much smaller for the case with pitch control. Of interest is that the rotor tuned for an intermediate $\Lambda/C_{f0}$ (such as Rotor-5 in this case) performs fairly well for a wide range of $\Lambda/C_{f0}$, suggesting that such an intermediate design would be preferred when the rotors are planned to be used in multiple farms with different effective farm densities.

## 4. Concluding remarks

In this paper, we have presented a novel theoretical analysis method called 'BEM-FM' to explore some potentially important relationships between the design of (horizontal-axis) wind turbine rotors and their performance in a very large wind farm. The BEM-FM model presented in this paper is the first model of its kind and could be extended further in future studies; hence the example solutions presented are primarily for the purpose of demonstrating how this new method works. Nonetheless, these theoretical results suggest that the benefits of tuning rotor design for a given farm density could be substantial and are worthy of further consideration. In particular, it has been shown that the peak power coefficient of rotors 'tuned' for a given farm density could be significantly higher than that of 'untuned' rotors if the blade pitch angle is not allowed to be adjusted during operation, and could still be noticeably higher even if the blade pitch angle is allowed to be adjusted optimally. These results essentially support the concept of 'Level 3' optimisation of large wind farms described earlier in this paper, i.e., the overall efficiency of large wind farms could be improved by optimising (or tuning) the design of turbines as well as their arrangements and operating conditions.

To focus on the fundamentals of the two-scale coupled nature of this problem, the optimisation of turbine rotor design presented in this paper has been simplified significantly, considering only the aerodynamic performance (or power coefficient) of ideal three-bladed rotors as the objective to be maximised. In the real world of turbine rotor design, it is essential to consider some other important factors, such as the structural performance (or durability) of turbine blades and the necessity of power capping strategies (to protect the whole turbine system including the generator and gearbox when the wind speed exceeds the rated wind speed), thereby reducing the levelised cost of electricity (LCOE). To address such industrial challenges, the optimisation process presented in this paper would need to be modified in future studies (for example, the dimensional farm-layer wind speed $U_F$ would need to be considered explicitly in the optimisation process in addition to the non-dimensional wind speed reduction factors $\alpha$ and $\beta$). Nevertheless, it is worth noting that the classical BEM theory still plays a central role in the design of real wind turbine rotors of today. We therefore expect that the coupled BEM-FM method proposed here may become the basis of many applications in the future.

Finally, it should be remembered that the theoretical model presented in this paper is for ideal very large wind farms. To predict the performance of a real wind farm, some corrections would need to be applied to the current model to account for the influence of atmospheric and geographical conditions as well as the actual size of the farm. In particular, unless the farm size is as large as the scale of the relevant atmospheric system that drives the flow over the farm, the left-hand side of the farm-scale momentum balance equation (2.2) may in reality be larger than the right-hand side of (2.2), meaning that the current model may under-predict the power of a real farm. Therefore, an additional 'larger-scale' (larger than the farm scale) model describing the driving force of the flow over the farm in a more general manner would need to be coupled with the current theoretical model to generalise the present work. This practically important point needs to be investigated further in future studies. A preliminary analysis and discussion on this point are presented in Appendix B.

**Data accessibility.** The work presented in this paper is theoretical. Results can be reproduced from the information provided in the paper.

**Authors' contribution.** T.N. developed the theoretical model and drafted the entire manuscript. W.H. created the airfoil performance curve and helped revise the manuscript. Both authors contributed to the calculation and analysis of theoretical results and gave final approval for publication.

**Competing interests.** We have no competing interests.

**Funding.** There was no funding for the work presented in this paper.

**Acknowledgements.** T.N. would like to thank Dr Pierre-Luc Delafin, Dr Thomas Dunstan and Mr Edgar Perez-Campos for helpful discussions.

## Appendix A.  Analytical solutions of the two-scale momentum model

The original two-scale momentum model proposed by Nishino (2016) may be solved analytically for some special cases. Perhaps the most useful analytical solutions are those for $\gamma = 2$; in this case, the two-scale momentum equation (2.13) can be solved analytically to obtain $\beta$ as function of $\alpha$, and eventually, $C_P$ as a function of $\alpha$ (or alternatively $C_T'$) as follows:

$$C_P = 4\alpha^2(1-\alpha)\left(4\alpha(1-\alpha)\frac{\Lambda}{C_{f0}} + 1\right)^{-\frac{3}{2}} \tag{A.1}$$

$$= \frac{64C_T'}{(C_T'+4)^3}\left(\frac{16C_T'}{(C_T'+4)^2}\frac{\Lambda}{C_{f0}} + 1\right)^{-\frac{3}{2}}. \tag{A.2}$$

To obtain the optimal $\alpha$ value that maximises $C_P$ for a given $\Lambda/C_{f0}$, we may differentiate $C_P$ in (A.1) with respect to $\alpha$ and consider $\mathrm{d}C_P/\mathrm{d}\alpha = 0$, namely

$$\alpha^2 + \left(\frac{3}{2}\frac{C_{f0}}{\Lambda} - 1\right)\alpha - \frac{C_{f0}}{\Lambda} = 0. \tag{A.3}$$

By solving (A.3), we may obtain the optimal $\alpha$ value as

$$\alpha_{\text{opt}} = \left(1 - \frac{3}{2}\frac{C_{f0}}{\Lambda} + \sqrt{\left(\frac{3}{2}\frac{C_{f0}}{\Lambda} - 1\right)^2 + 4\frac{C_{f0}}{\Lambda}}\right)/2 \tag{A.4}$$

and hence, from (2.15), we may also obtain the optimal $C_T'$ value as

$$C_{T\text{opt}}' = \frac{4(1-\alpha_{\text{opt}})}{\alpha_{\text{opt}}}. \tag{A.5}$$

It is also possible (but more difficult) to obtain $C'_{Topt}$ by differentiating $C_P$ in (A.2) with respect to $C'_T$ and then finding an appropriate solution of $dC_P/dC'_T = 0$, namely

$$C'^3_T + \left(4\frac{\Lambda}{c_{f0}} + 6\right)C'^2_T + 16\frac{\Lambda}{c_{f0}}C'_T - 32 = 0 \ . \tag{A.6}$$

Finally, the maximum $C_P$ (corresponding to $C_{Pmax}$ plotted earlier in figures 2 and 6) can be obtained analytically by substituting $\alpha_{opt}$ into (A.1), or alternatively, $C'_{Topt}$ into (A.2).

## Appendix B. Generalisation of the two-scale momentum model

The original two-scale momentum model of Nishino (2016) is for an ideal very large wind farm, which is so large (as large as the scale of the relevant atmospheric system driving the flow over the farm) that the momentum balance equation (2.2) is approximately satisfied, i.e., the momentum loss (or pressure drop) from the most-upstream position to the most-downstream position of the farm site is approximately unchanged before and after farm construction. For a real wind farm that is large enough to assume that the flow through the farm is mostly fully developed (this is likely to be of the order of 10km) but not as large as the scale of the relevant atmospheric system (that drives the flow over the farm for a certain amount of time; this could be of the order of 100km or larger, depending on the wind generation mechanism), we need to consider that the aforementioned momentum loss could become larger after farm construction than before, since the farm could induce an additional pressure difference between the upstream and downstream sides of the farm (in a somewhat similar manner to how a turbine rotor induces a pressure difference between the upstream and downstream sides of it). Hence, the momentum balance equation (2.2) needs to be replaced by

$$\langle \tau_w \rangle S + T = \frac{\Delta p}{\Delta p_0} \tau_{w0} S \ , \tag{B.1}$$

where $\Delta p_0$ and $\Delta p$ denote the streamwise pressure drop observed before and after farm construction, respectively (note that here the pressure is considered to vary only linearly within the farm site since the flow through the farm site has been assumed to be fully developed, not only before but also after farm construction). Eventually, (2.10) will need to be replaced by

$$C^*_T \frac{\Lambda}{c_{f0}}\beta^2 + \beta^\gamma - 1 = \frac{\Delta p - \Delta p_0}{\Delta p_0}, \tag{B.2}$$

and (2.13) by

$$4\alpha(1-\alpha)\frac{\Lambda}{c_{f0}}\beta^2 + \beta^\gamma - 1 = \frac{\Delta p - \Delta p_0}{\Delta p_0} \ . \tag{B.3}$$

The only difference of the above 'generalised' two-scale momentum equation (B.3) from the original equation (2.13) is in the right-hand side, which now takes a non-zero value unless $\Delta p = \Delta p_0$. In other words, the original equation (2.13) is for a special case where $\Delta p = \Delta p_0$.

The value of the right-hand side of (B.3) is expected to depend largely on $\beta$ as well as on the ratio of the farm size to the size of the relevant atmospheric system driving the flow. However, this may also depend on other atmospheric and geographical conditions that are site-specific and is therefore difficult to be estimated purely theoretically. To obtain a quantitative estimate of the right-hand side of (B.3) for a given large wind farm, it would be necessary to conduct some numerical simulations of atmospheric flow over the given farm site. For further analysis and discussion on this issue, readers are referred to a recent paper by Nishino (2018).

REFERENCES

ADAMS, A. S. & KEITH, D. W. 2013 Are global wind power resource estimates overstated? *Environ. Res. Lett.* **8**, 015021.

ARCHER, C. L., MIRZAEISEFAT, S. & LEE, S. 2013 Quantifying the sensitivity of wind farm performance to array layout options using large-eddy simulation. *Geophys. Res. Lett.* **40**, 4963-4970.

CALAF, M., MENEVEAU, C. & MEYERS, J. 2010 Large eddy simulation study of fully developed wind-turbine array boundary layers. *Phys. Fluids* **22**, 015110.

CHAMORRO, L. P., TOBIN, N., ARNDT, R. E. A. & SOTIROPOULOS, F. 2014 Variable-sized wind turbines are a possibility for wind farm optimization. *Wind Energ.* **17**, 1483-1494.

DRELA, M. 1989 XFOIL: an analysis and design system for low Reynolds number airfoils, in T. J. Mueller (ed.) *Low Reynolds Number Aerodynamics*, Springer, pp. 1-12.

DUNSTAN, T., MURAI, T. & NISHINO, T. 2018 Validation of a theoretical model for large turbine array performance under realistic atmospheric conditions. *AMS 23rd Symposium on Boundary Layers and Turbulence*, 11-15 June, Oklahoma City, OK, USA, Paper 13A.3.

EMEIS, S. & FRANDSEN, S. 1993 Reduction of horizontal wind speed in a boundary layer with obstacles. *Bound.-Layer Meteor.* **64**, 297-305.

FLEMING, P. A., GEBRAAD, P. M. O., LEE, S., VAN WINGERDEN, J.-W., JOHNSON, K., CHURCHFIELD, M., MICHALAKES, J., SPALART, P. & MORIARTY, P. 2014 Evaluating techniques for redirecting turbine wakes using SOWFA. *Renew. Energ.* **70**, 211-218.

FRANDSEN, S. 1992 On the wind speed reduction in the center of large clusters of wind turbines. *J. Wind Eng. Ind. Aerodyn.* **39**, 251-265.

GHAISAS, N. S., GHATE, A. S. & LELE, S. K. 2017 Farm efficiency of large wind farms: evaluation using large eddy simulation. *Proc. Tenth International Symposium on Turbulence and Shear Flow Phenomena*, 6-9 July, Chicago, IL, USA, 6pp.

GLAUERT, H. 1935 Airplane propellers, in W. F. Durand (ed.) *Aerodynamic theory*, Springer, pp. 169-360.

HANSEN, M. L. O. 2015 *Aerodynamics of wind turbines*, Routledge.

MA, L. & NISHINO, T. 2018 Preliminary estimate of the impact of support structures on the aerodynamic performance of very large wind farms. *J. Phys.: Conf. Ser.* **1037**, 072036.

MEYERS, J. & MENEVEAU, C. 2012 Optimal turbine spacing in fully developed wind farm boundary layers. *Wind Energ.* **15**, 305-317.

MILLER, L. M., BRUNSELL, N. A., MECHEM, D. B., GANS, F., MONAGHAN, A. J., VAUTARD, R., KEITH, D. W. & KLEIDON, A. 2015 Two methods for estimating limits to large-scale wind power generation. *PNAS* **112**, 11169-11174.

NING, S. A. 2013 AirfoilPrep.py documentation release 0.1.0. Technical Report NREL/TP-5000-58817, National Renewable Energy Laboratory, USA.

NISHINO, T. 2016 Two-scale momentum theory for very large wind farms. *J. Phys.: Conf. Ser.* **753**, 032054.

NISHINO, T. 2018 Generalisation of the two-scale momentum theory for coupled wind turbine/farm optimisation. *25th National Symposium on Wind Engineering*, 3-5 December, Tokyo, Japan, 6pp. (accepted for publication)

NISHINO, T. & WILLDEN, R. H. J. 2013 Two-scale dynamics of flow past a partial cross-stream array of tidal turbines, *J. Fluid Mech.*, **730**, 220-244.

STEVENS, R. J. A. M., GAYME, D. F. & MENEVEAU, C. 2014 Large eddy simulation studies of the effects of alignment and wind farm length. *J. Renew. Sustain. Energy* **6**, 023105.

STEVENS, R. J. A. M., GAYME, D. F. & MENEVEAU, C. 2016 Effects of turbine spacing on the power output of extended wind-farms. *Wind Energ.* **19**, 359-370.

VAN KUIK, G. A. M., PEINKE, J. AND 25 OTHERS 2016 Long-term research challenges in wind energy – a research agenda by the European Academy of Wind Energy. *Wind Energ. Sci.* **1**, 1-39.

WHELAN, J. I., GRAHAM, J. M. R. & PEIRÓ, J. 2009 A free-surface and blockage correction for tidal turbines, *J. Fluid Mech.* **624**, 281-291.

XIE, S., ARCHER, C. L., GHAISAS, N. & MENEVEAU, C. 2017 Benefits of collocating vertical-axis and horizontal-axis wind turbines in large wind farms. *Wind Energ.* **20**, 45-62.

WU, Y.-T. & PORTÉ-AGEL, F. 2013 Simulation of turbulent flow inside and above wind farms: model validation and layout effects. *Bound.-Layer Meteor.* **146**, 181-205.

ZAPATA, A., NISHINO, T. & DELAFIN, P.-L. 2017 Theoretically optimal turbine resistance in very large wind farms. *J. Phys.: Conf. Ser.* **854**, 012051.